\documentclass{article}

\usepackage[bookmarks = true, colorlinks=true, linkcolor = blue, citecolor = blue, menucolor = blue, urlcolor = blue]{hyperref}
\usepackage{graphicx}
\usepackage{amssymb,amsmath}
\usepackage{color}
\usepackage{listings}
\usepackage{tikz}
\usepackage{verbatim}
\usetikzlibrary{decorations.markings}
\usetikzlibrary{arrows}
\usepackage{theorem}
\usepackage{fancyvrb}
\usepackage{cite}
\usepackage{rotating}
\newtheorem{Thm}{Theorem}

\newtheorem{example}[Thm]{Example}
\usepackage{calc}

\newcommand{\C}[1]{\mbox{\lstinline`#1`}}

\definecolor{dkblue}{rgb}{0,0.1,0.5} 
\definecolor{lightblue}{rgb}{0,0.5,0.5} 
\definecolor{dkgreen}{rgb}{0,0.4,0} 
\definecolor{dk2green}{rgb}{0.4,0,0} 
\definecolor{dkviolet}{rgb}{0.6,0,0.8}
\definecolor{shadethmcolor}{rgb}{0.9, 0.9,1}

\author{J\'onathan Heras and Ekaterina Komendantskaya}

\title{HoTT formalisation in Coq: Dependency Graphs \& ML4PG}

\begin{document} 

\maketitle

\section{Introduction}

This note is a response to Bas Spitter's email of 28 February 2014 about ML4PG:

\emph{We (Jason actually) are adding dependency graphs to our HoTT library:\\
\url{https://github.com/HoTT/HoTT/wiki}}\\

\emph{
I seem to recall that finding the dependency graph was a main obstacle preventing machine learning to be used in Coq. However, you seem to have made progress on this. What tool are you using?
\url{https://anne.pacalet.fr/dev/dpdgraph/} ?}\\

\emph{Or another tool?  Would it be easy to use your ML4PG on the HoTT library?}

This note gives explanation of how ML4PG can be used in the HoTT library and how ML4PG relates to the two kinds of Dependency graphs available in Coq.

\section{Dependency Graphs in Coq: }

There are two kinds of dependency graphs in Coq:

\begin{enumerate}
	\item[\textbf{DG-1}] \emph{Graphs showing dependency of a Coq theorem to all the auxiliary results that were used to prove that theorem}. These graphs 
	can be generated using the \emph{dpdgraph} tool available at~\cite{dpdgraph}. This tool can be also used to generate the dependencies of all the theorems of a library (see~\url{http://hott.github.io/HoTT/file-dep-graphs/HoTT.types.Paths.svg}).

\begin{example}\label{ex1}
In the \verb"Paths" file of the Hott library~\cite{hottbook}, we can find the following two theorems.

\begin{lstlisting}
Lemma dpath_path_l {A : Type} {x1 x2 y : A}
  (p : x1 = x2) (q : x1 = y) (r : x2 = y) :
  q = p @ r <~> transport (fun x => x = y) p q = r.
Proof.
  destruct p; simpl.
  exact (equiv_concat_r (concat_1p r) q).
Qed.

Lemma dpath_path_lr {A : Type} {x1 x2 : A}
  (p : x1 = x2) (q : x1 = x1) (r : x2 = x2) :
  q @ p = p @ r <~> transport (fun x => x = x) p q = r.
Proof.
  destruct p; simpl.
  refine (equiv_compose' (B := (q @ 1 = r)) _ _).
  exact (equiv_concat_l (concat_p1 q)^ r).
  exact (equiv_concat_r (concat_1p r) (q @ 1)).
Qed.
\end{lstlisting}

The dependency graphs, generated with the \emph{dpdgraph} tool, of these two lemmas are given in Figures~\ref{dgraph1} and~\ref{dgraph2}, respectively.

\begin{figure}
\centering
\includegraphics[scale=.25]{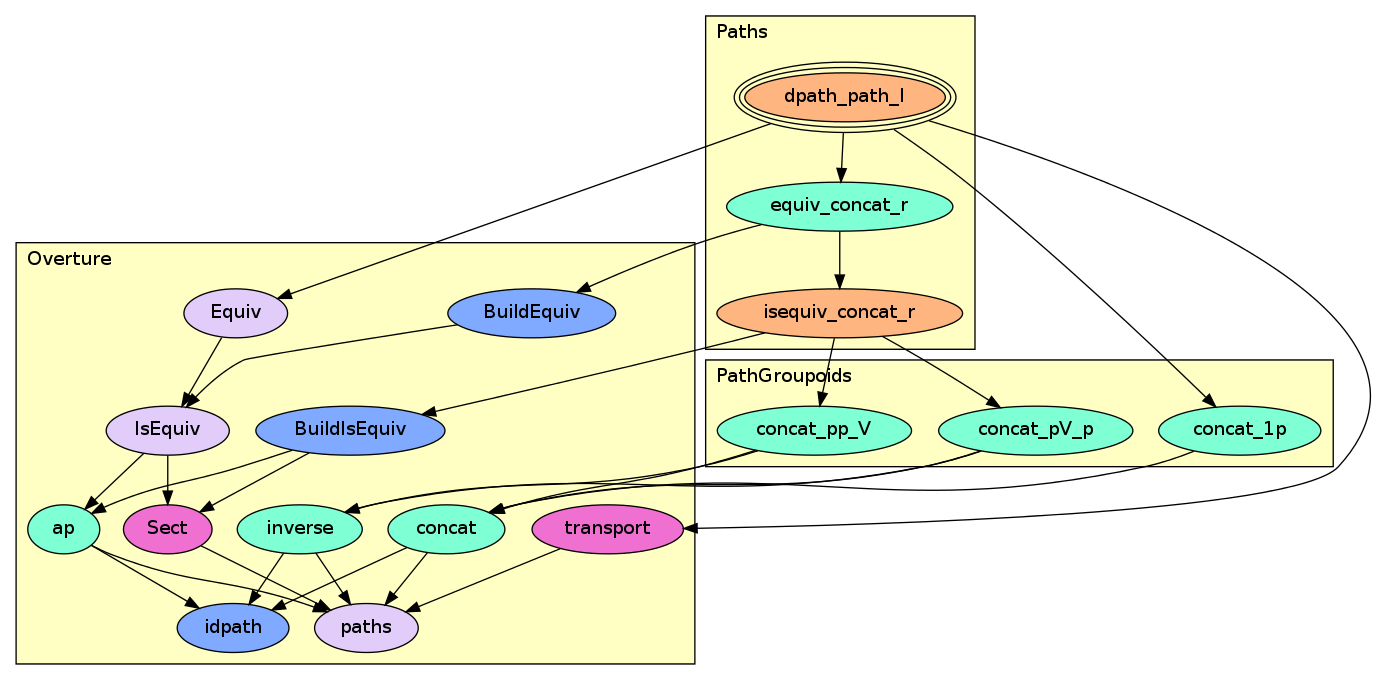}
\caption{\emph{Dependency graph for the theorem \texttt{dpath\_path\_l} included in the \texttt{Paths} library of HoTT.}}\label{dgraph1}
\end{figure}

\begin{figure}
\centering
\includegraphics[scale=.15]{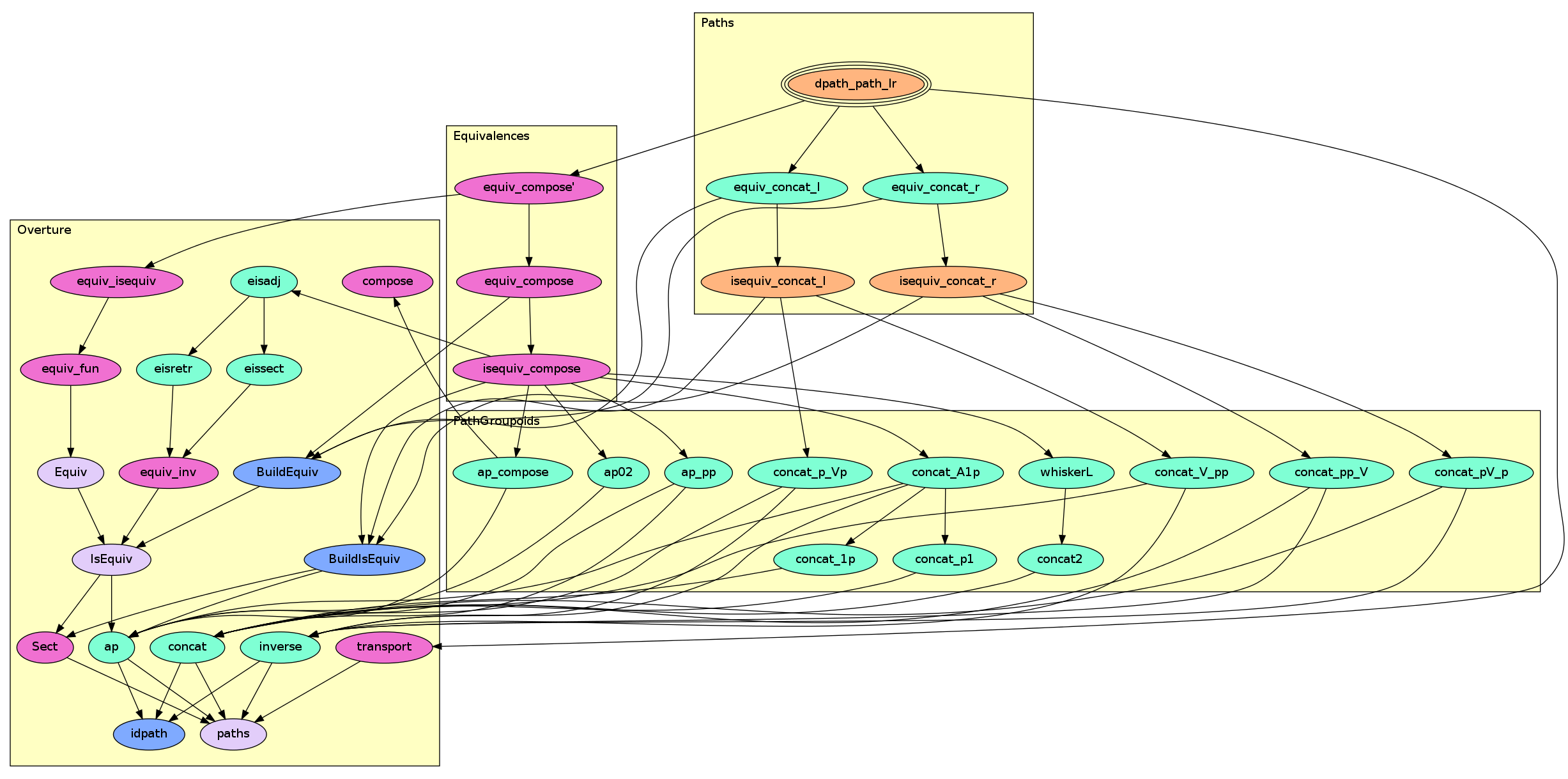}
\caption{\emph{Dependency graph for the theorem \texttt{dpath\_path\_lr} included in the \texttt{Paths} library of HoTT.}}\label{dgraph2}
\end{figure}

\end{example}

	The graphs generated with the \emph{dpdgraph} tool should be read as follows.
	
	\begin{itemize}
	 \item A box represents a library file. E.g. Figure~\ref{dgraph1} shows that the proof of \verb"dpath_path_l" involves results from the \verb"Paths", \verb"PathGroupoids" and \verb"Overture" libraries.
	 \item An edge $n1 \rightarrow n2$ means that $n1$ uses $n2$. If $n1$ uses $n2$ and $n3$, and $n2$ uses $n3$; then, the only edges that appear in the graph are  $n1\rightarrow n2$ and $n2 \rightarrow n3$.
	 \item A multi-circled node $n$ indicates that $n$ is not used (no predecessor in the graph) -- if we generate the dependencies for a theorem $t$, the node for $t$ will be the only multi-circled node. 
	 \item Orange nodes are theorems 
	 \item Green nodes are definitions.
	 \item Light pink nodes are classes, or Inductive types.
	 \item Blue nodes are constructors of the classes or Inductive types.
	 \item Dark pink nodes are constructors inside the constructors of classes.
	\end{itemize}

	\item[\textbf{DG-2}] \emph{Graphs showing relations between libraries}. In these graphs, nodes represent library files, and an edge  $l_1 \rightarrow l_2$ means that $l_2$ was imported in $l_1$. The graph is shown after a transitive reduction process. This kind of graphs can be generated using scripts from different projects like the MathComponents library~\cite{FTT} or the MathClasses library~\cite{mathclasses}. Some examples of this kind of dependency graphs are available at:
	
	\begin{itemize}
	 \item \url{http://hott.github.io/HoTT/dependencies/HoTTCore.svg}.
	 \item \url{http://ssr.msr-inria.inria.fr/~jenkins/current/index.html}.
	 \item \url{http://www.unirioja.es/cu/cedomin/crship/BPL_general/toc.html}.
	\end{itemize}

\end{enumerate}

%
%

In addition to the two above mentioned methods to generate dependency graphs, a method to extract dependencies from 
Coq libraries was presented in~\cite{urbancoq}; however, to the best of our knowledge, this method does not generate any 
graphical representation for the captured dependencies.

\section{ML4PG: how it works on the example of the HoTT library}

ML4PG~\cite{KHG13,HK14,HK14mcs} is a machine-learning extension to the Proof General interface for Coq. It works in the background of Proof General capturing statistical features from
Coq definitions and theorems. On user's demand, the statistical features are sent to a machine-learning tool to find families of similar definitions, theorems or proofs using clustering algorithms (a family of unsupervised machine-learning algorithms that groups together families of similar objects in \emph{clusters}).

ML4PG implements two different methods to extract features from Coq terms and Coq proofs respectively, see~\cite{HK14} for a detailed explanation of the two feature-extraction algorithms.

\textbf{(*)} The first feature-extraction algorithm collects statistical features from terms. In particular, given a term $t$, we generate its ML4PG-term-tree (cf. Figure~\ref{ml4pgtermtree}) and from this term-tree we generate a feature-table that captures the structure of the term-tree and the terms and types at each node of the tree (cf. Figure~\ref{ml4pgtermtree}).

%
%
%
%

\begin{example}

Given the term of the statement of Lemma \verb"dpath_path_l", its ML4PG-term-tree and (a fragment of its) feature-table are given in
Figure~\ref{ml4pgtermtree}.

\begin{sidewaysfigure}[!]
\centering
\includegraphics[scale=.175]{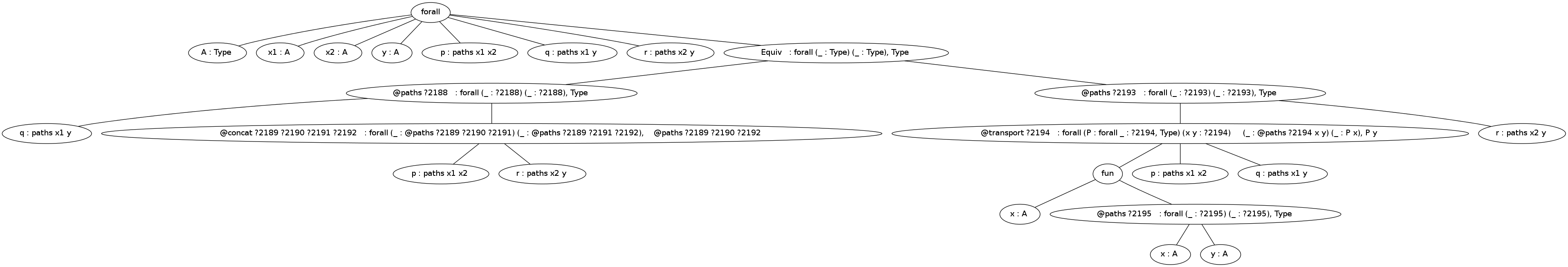}
{\scriptsize 
\begin{tabular}{|c||c|c|c|c|c|c|}
\hline
 &  level index 0 & level index 1 & level index 2  & level index 3  & level index 4  & li 5   \\
 \hline
td0 & ($[\texttt{forall}]_{Gallina}$,-1,-1) & (0,0,0) & (0,0,0) & (0,0,0) & (0,0,0) & \ldots \\
\hline
td1 & ($[\texttt{A}]_{term}$,$[\texttt{Type}]_{type}$,0) & ($[\texttt{x1}]_{term}$,$[\texttt{A}]_{type}$,0)& ($[\texttt{x2}]_{term}$,$[\texttt{A}]_{type}$,0) & ($[\texttt{y}]_{term}$,$[\texttt{A}]_{type}$,0) & ($[\texttt{p}]_{term}$,$[\texttt{paths x1 x2}]_{type}$,0) & \ldots \\
\hline
td2 & ($[\texttt{paths ?X}]_{term}$,$[\texttt{?X -> ?X -> Type}]_{type}$,7) & ($[\texttt{paths ?Y}]_{term}$,$[\texttt{?Y -> ?Y -> Type}]_{type}$,7) & (0,0,0) & (0,0,0) &(0,0,0) & \ldots\\
\hline

\end{tabular}}
 {\scriptsize 
\begin{tabular}{|c||c|c|c|c|}
\hline
 &  li 4  & level index 5  & level index 6  & level index 7 \\
 \hline
td0 & \ldots & (0,0,0) & (0,0,0) & (0,0,0) \\
\hline
td1 & \ldots & ($[\texttt{q}]_{term}$,$[\texttt{paths x1 y}]_{type}$,0) & ($[\texttt{r}]_{term}$,$[\texttt{paths x2 y}]_{type}$,0) & ($[\texttt{Equiv}]_{term}$,$[\texttt{Type -> Type -> Type}]_{type}$,0)\\
\hline
td2 & \ldots  & (0,0,0) & (0,0,0) & (0,0,0)\\
\hline
\end{tabular}}
\caption{\emph{ML4PG-term-tree and a fragment of the term-feature-table for the term of the statement of Lemma \texttt{dpath\_path\_l}.} The rows of the term-feature-table represent the depth of the tree, and the columns the index level. The element $(i,j)$ of the table encodes the node of depth $i$ and index-level $j$ (e.g. the element $(1,4)$ of the table encodes the node \texttt{p : paths x1 x2} of the term tree. Each node is represented by a triple that captures its term component, its type component and its parent node.}
\label{ml4pgtermtree}
\end{sidewaysfigure}

\end{example}

The process of feature extraction mimics very closely the dependency graph for Coq terms. To populate the cells of the feature extraction table with feature values, the ML4PG tracks the dependencies of the auxiliary terms used
within the term in question. For example, the term-tree for   Lemma \texttt{dpath\_path\_l} (see Figure~\ref{ml4pgtermtree}) contains a sub-expression \texttt{p : paths x1 x2}, and to extract a feature table for Lemma \texttt{dpath\_path\_l} one must compute the representative numeric value for \texttt{p : paths x1 x2}. To compute the latter, we must cluster  \texttt{p : paths x1 x2} against other, previously defined terms and types, and they themselves must have feature tables, and so on, the process continues recursively until it reaches the basic library definitions, and includes dependencies of multiple theorems, definitions, lemmas etc. in the statistical pattern-recognition.
In~\cite{KHG13,HK14}, we call it \emph{recurrent clustering}.


The construction of the term-feature-tables is a process that runs in the background of ML4PG. When the user asks ML4PG to show families of similar definitions or lemma statements, ML4PG processes the tables, sends them to a clustering algorithm implemented in Weka~\cite{Weka} (a machine-learning interface), and when the clustering algorithm is completed, ML4PG post-processes the results for showing them to the user.

\begin{figure}[!]
\centering
\includegraphics[scale=.3]{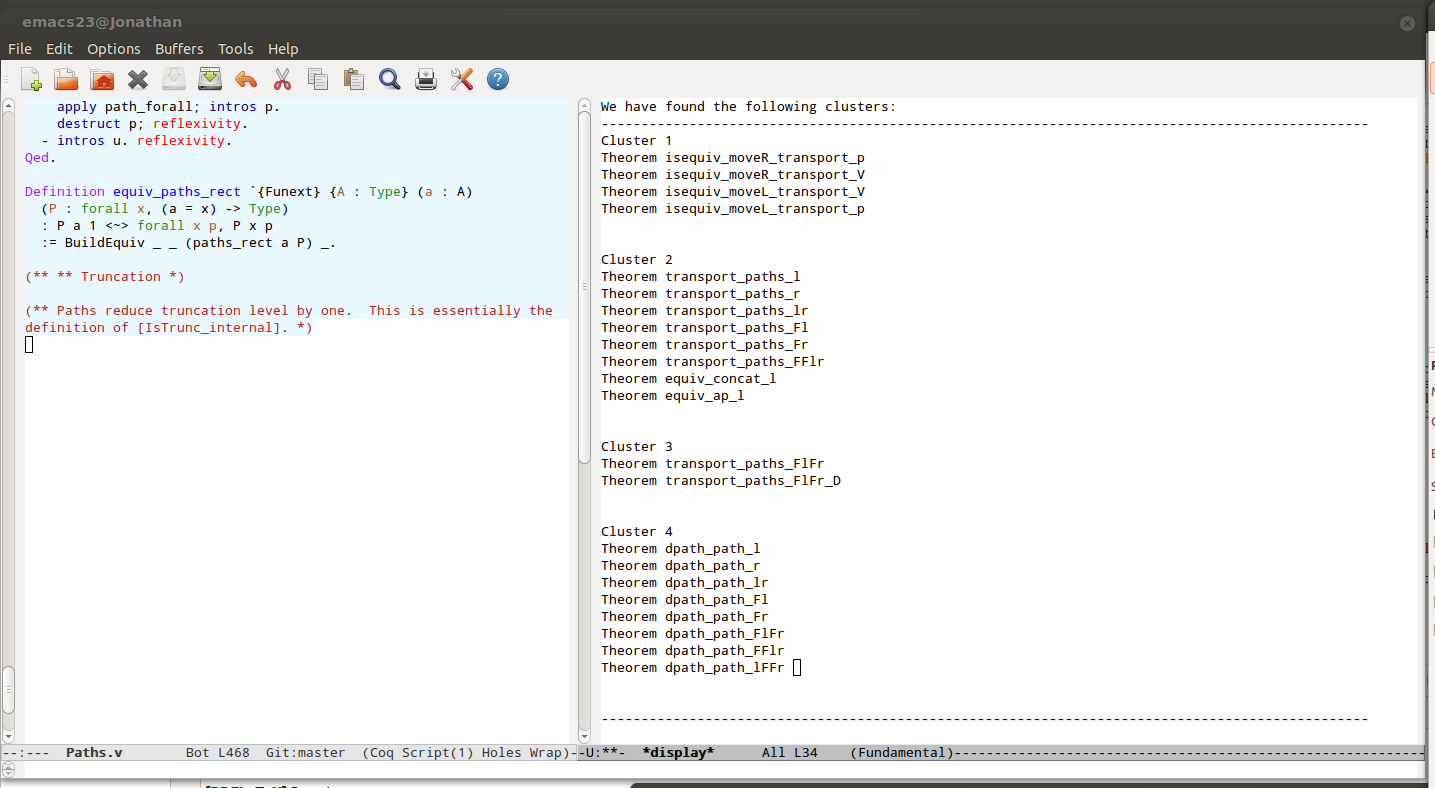}
\caption{\emph{ML4PG's output for term-clustering of lemmas of the Paths library using granularity value 3.} The Proof General screen has been split into two buffers. The left buffer shows the current proof-development (in this case, the \texttt{Paths} library of HoTT); and, the right buffer shows the families of similar lemma statements.}\label{ml4pgoutput} 
\end{figure}

\begin{example}

ML4PG can be used to cluster the lemma statements of the Paths library of HoTT. The families of similar lemma statements are shown in Figure~\ref{clustergraph2} --- the current output of ML4PG is displayed in a buffer of Proof General using plain text with links to the lemmas (see Figure~\ref{ml4pgoutput}), the similarity graph of Figure~\ref{clustergraph2} can be generated on user's demand. ML4PG differentiates without any user interaction lemmas about transporting in path spaces (lemmas with the prefix \verb"transport"), the move family of equivalences (lemmas with the prefix \verb"isequiv_move"), dependent paths (lemmas with the prefix \verb"dpath_path"). Among these families, ML4PG also discovers lemma statements that are closer among them. 

\begin{figure}[!]
\centering
\includegraphics[scale=.23]{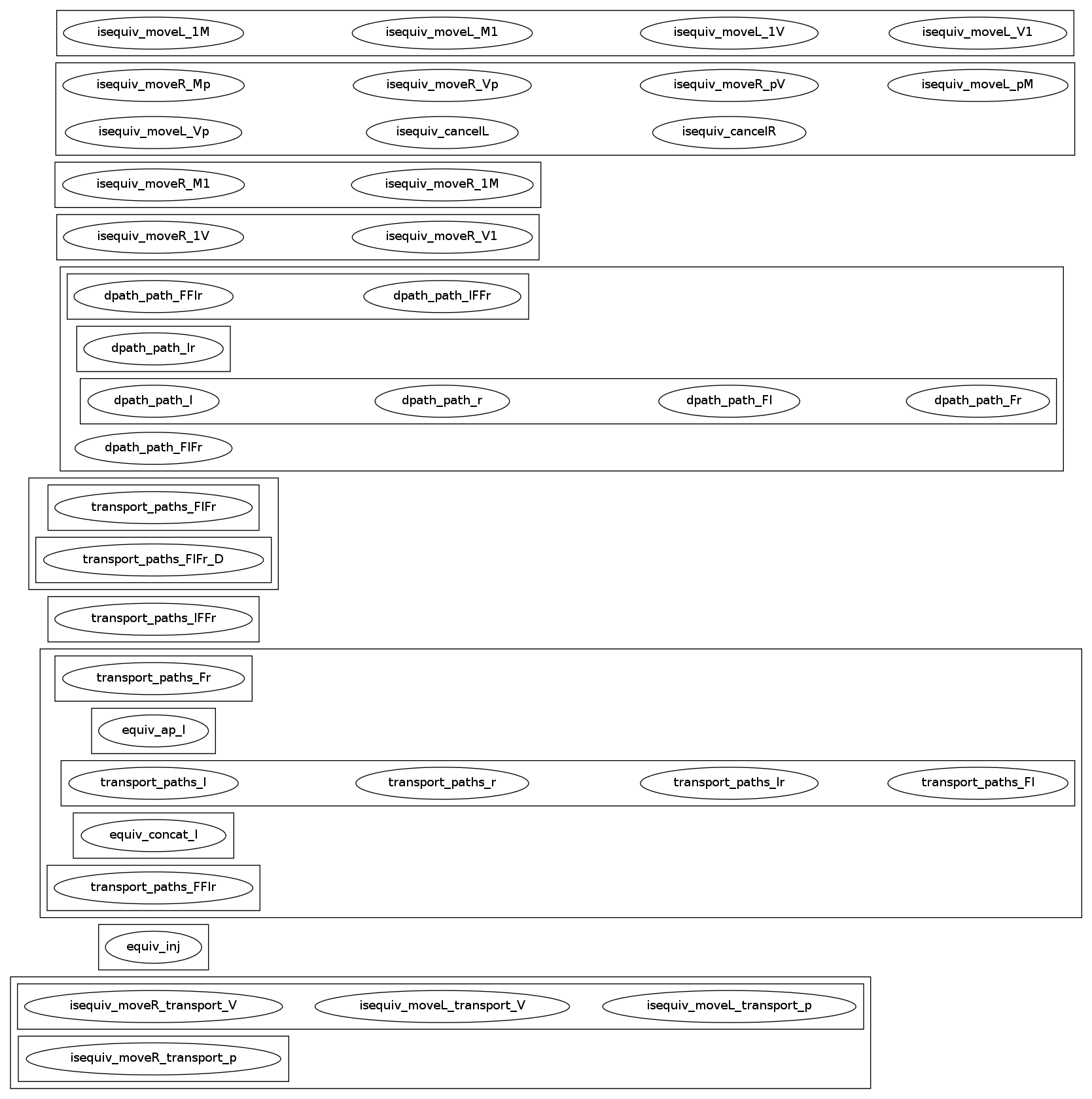}
\caption{\emph{Visualisation of the term-clusters of the Paths library.} The boxes of the diagram represent the different clusters, and the ellipses inside the boxes are the lemmas of each cluster. The precision of clusters can be adjusted using the granularity value -- an ML4PG parameter (whose value is between 1 and 5) to adjust the precision of the clusters (1=low precision, 5=high precision)~\cite{KHG13} -- in particular, the clusters obtained with a low granularity value are usually split into smaller clusters when the granularity value is increased. We run twice the clustering algorithm to generate the visualisation of term-clusters: first with granularity value $3$ and then with granularity value $5$. If a cluster that appears with granularity value $3$ is split in smaller clusters when increasing the granularity value, we represent this situation using boxes inside boxes (e.g. Lemmas \texttt{isequiv\_moveR\_transport\_V}, \texttt{isequiv\_moveL\_transport\_V}, \texttt{isequiv\_moveL\_transport\_p} and \texttt{isequiv\_moveR\_transport\_p} are 
grouped together with granularity $3$; however, when increasing the granularity value, only the first three lemmas are grouped together, cf. the last cluster of the figure). }\label{clustergraph2}
\end{figure}
\end{example}

Note that the dependencies of a term are captured with our feature-extraction method. Given a term $t$, the feature table of $t$ stores the terms that form $t$; and, in turn, the dependencies of those terms are captured thanks to the recurrent clustering process.

The above mentioned method can cluster similar statements of all Coq terms, including lemmas and theorems. However, this method does not capture the interactive nature of Coq proofs. To solve this problem, we have a different method to extract features from Coq proofs, see~\cite{HK14}. 

\textbf{(**)} The method captures some important properties present in proof subgoals: tactics applied, type of the arguments of the tactics, lemmas applied, top symbols of the goal, and number of subgoals. These features are stored into numerical feature-tables that will be processed by machine-learning algorithms. To transform features like types, lemmas applied, or top symbols we again use recurrent term-clustering to assign similar values to objects that belong to the same cluster.

\begin{example}
The feature table for the proof of Lemma \verb"dpath_path_l" (cf. Example~\ref{ex1}) is given in Table~\ref{prooftable}.

\begin{table}
\centering
\scriptsize{
\begin{tabular}{|l||l|l|l|l|l|l|}
\hline
 & \emph{tactics} & \emph{n} & \emph{arg type} & \emph{arg} & \emph{symbols} & \emph{goals} \\
\hline
\hline
\emph{g1}& $([destruct]_{tac},$ & $2$  & $([\verb"paths x1 x2"]_{type}$  & $(0,0,0,0)$ & $([\verb"<~>"]_{term},$ & $1$ \\
& $[simpl]_{tac},0,0)$ &   & $0,0,0)$  &  & $[=]_{term},[=]_{term})$ &  \\
 \hline
  \emph{g2} & $([exact]_{tac},$ & $1$  & $([Prop]_{type},0,0,0)$  & $([(\verb"equiv_concat_r (concat_1p r) q")]_{term},$ & $([\verb"<~>"]_{term},$& $0$ \\
 & $0,0,0)$ &  &  & $0,0,0)$ & $[=]_{term},[=]_{term})$& \\
 
  \hline
  \end{tabular}}
\caption{\emph{Feature table for the proof of Lemma \texttt{dpath\_path\_l}.}}\label{prooftable}
\end{table}

\end{example}

The feature-tables obtained from the proofs of a library (or several libraries) are sent on user's demand to clustering algorithms to find families of similar proofs. 

\begin{figure}
\centering
\includegraphics[scale=.2]{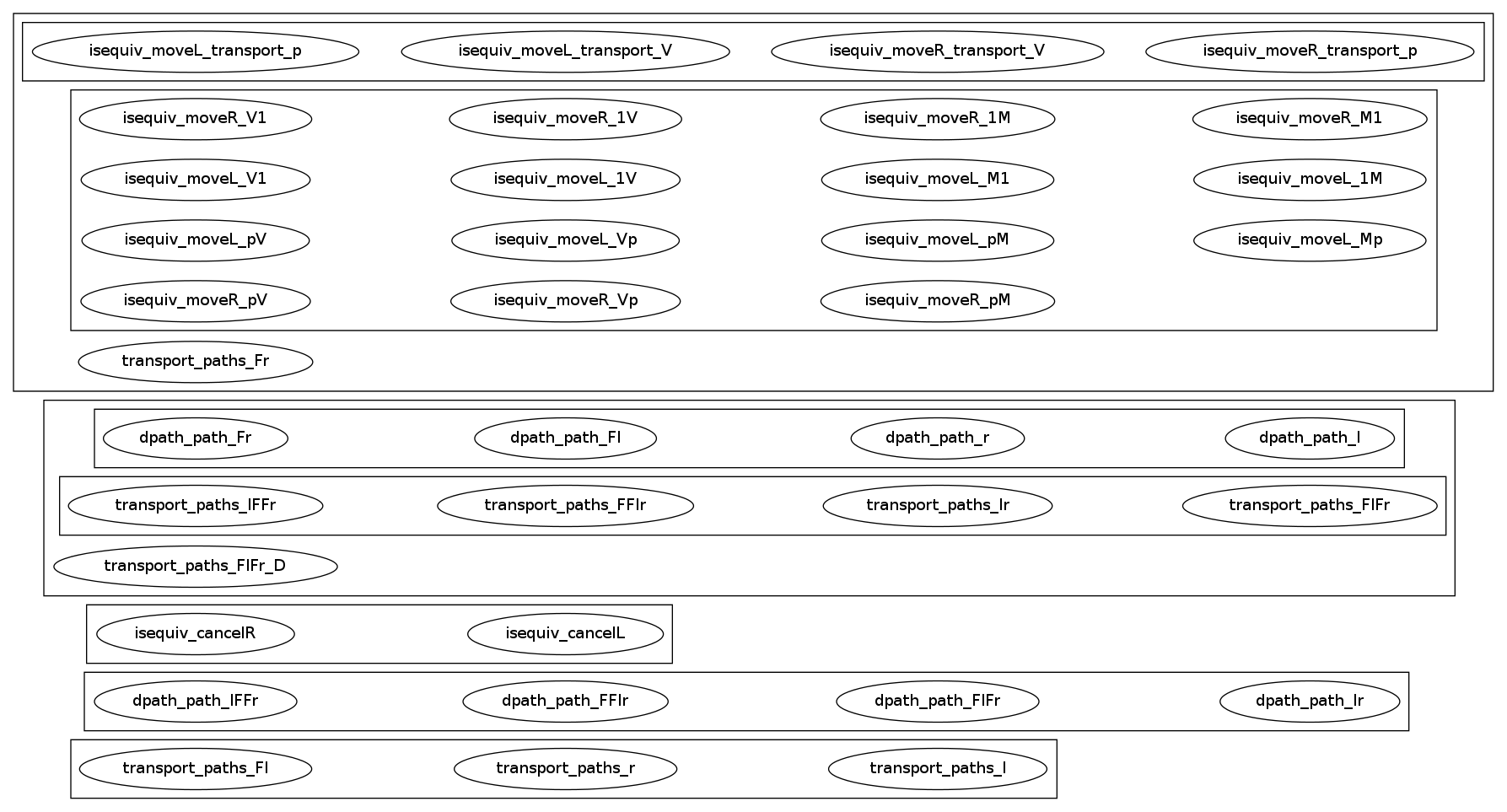}
\caption{\emph{Visualisation of the proof-clusters of the Paths library; based on analysis of their term structure and proofs.}}\label{clustergraph1}
\end{figure}

\begin{example}
ML4PG can be used to cluster the proof of the Paths library of HoTT. The families of similar proofs are shown in Figure~\ref{clustergraph1}. We can distinguish 5 proof strategies in this library:

\begin{itemize}
 \item A first group of proofs (\verb"dpath_path_lr", \verb"dpath_path_FlFr", \ldots) is proved using first case analysis in one path, then a refinement step; and the proofs finished using two \verb"exact" tactics.
 \item A second group of proofs (\verb"dpath_path_l", \verb"transport_paths_lr", \ldots) is proved  using first case analysis in one path and then the \verb"exact" tactic (with different arguments).
 \item A third group of proofs (\verb"transport_paths_r", \verb"transport_paths_l", \ldots) is proved using case analysis in two paths and then reflexivity.
 \item A fourth group of proofs (\verb"isequiv_cancelR" and \verb"isequiv_cancelL") is proved using case analysis in two paths, then simplification and finally the \verb"apply" tactic.
 \item The last group of proofs (\verb"isequiv_moveR_1V", \verb"isequiv_moveR_1M", \ldots) is proved using case analysis in one path and then the proofs are finished using the \verb"apply" tactic.
 \end{itemize}

\end{example}

As in the case of feature-extraction for terms, the feature-extraction for proofs capture the dependencies of proofs.

%

\section{Relation between the Dependency Graphs and ML4PG}

In this section, we explain the relation of ML4PG to dependency graphs. We divide this section into two parts -- as there are two kinds of dependency graphs present in Coq.

\subsection{Dependency graphs for the single lemma or term}

If we look into Figure~\ref{dgraph1}  of the dependency graph for Lemma \verb"dpath_path_l", we will see that:

\begin{enumerate}
\item it is designed to work with proofs rather than object definitions;
	\item it does not show the term structure \emph{per se}, but only the dependencies between the auxiliary lemmas/constructors used;
	\item it gives a complete information of all the results that were necessary to prove the lemma.
	\item extracting this information does not require any statistical machine-learning (just careful tracing of dependencies...)
	\item it may be hard to read, due to the presence of this complete information.
	\item hence, statistical machine learning may be useful for data-mining the above information in order to discriminate unimportant features of the lemma, and highlight those that are important.
\end{enumerate}

ML4PG essentially does what the last item says. The feature extraction of ML4PG known as ``recurrent  clustering'' (see~\cite{HK14}) takes into account the information represented in the dependency graph as follows:

\begin{itemize}
	\item ML4PG captures term and lemma dependencies recurrently, via the feature extraction which itself involves recurrent clustering of all previously defined terms;
	\item ML4PG's proof clustering in particular is very close to dependency graphs for lemmas, in that both are capturing mutual or inductive dependencies of all lemmas needed to prove the given statement.
\end{itemize}

As a result, ML4PG can be seen as a post-processing tool for dependency graphs. Given a big graph as shown in Figure~\ref{dgraph1}, what is the right way to discriminate its unimportant features? 
How to decide which of them are important or not?
One of the ways to do this, is to determine that relatively to other lemmas in the library. This is achieved by applying clustering in ML4PG.
Taking all of the above features into account, ML4PG can associate various object definitions and lemma statements and lemma proofs.
See Figures~\ref{clustergraph2} and~\ref{clustergraph1}.

\begin{figure}
\centering 
\includegraphics[scale=.25]{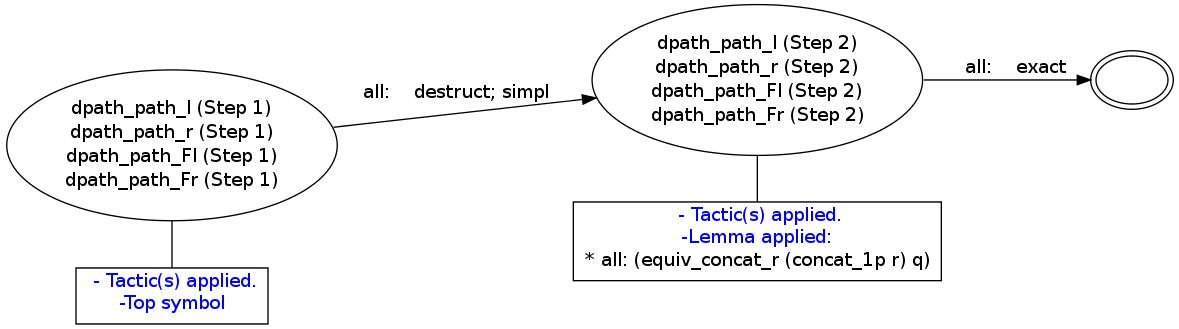} 
\caption{Automaton for one of the clusters of the Paths library using granularity value 5.}\label{automaton}
\end{figure}

On the basis of this association, we can essentially prune the statically generated dependency graphs to much smaller pictures, showing 
just some features of the given lemma and/or its proof, see Figures~\ref{automaton} and~\ref{automaton2}.

ML4PG's term-structure clustering goes beyond dependency graphs for terms, and carefully captures the tree term shape, structure, and dependency on other term- and type- structures present in the library (see Figure~\ref{ml4pgtermtree}):
\begin{itemize}

	\item The above approach makes sure that terms  appearing in the lemma (as well as their structures) are taken into account.
	\item Dependency of auxiliary terms on other terms defined in Coq libraries is taken into account recurrently.
	\item Types used in the statements of lemmas  are taken into account.
	\item If the types have been defined previously and depend on other types, all these hierarchies of dependencies are taken into account recurrently.
	\item Proof-clustering uses results of term-clustering via recurrent feature extraction.
	

	\item Unlike dependency graphs, tactics (not only auxiliary lemmas) used in the proof are taken into account.
\item Type structures of tactic arguments are taken into account. 
\end{itemize}

The work is under way to improve visualisation of the clustering results obtained by ML4PG.
So far, we have two main directions:

\begin{enumerate}
	\item Visualisation of the proof-flow.
	\item Visualisation of the synthesized (generalized) term-tree.
\end{enumerate}

The visualisation of the proof-flow summarising the important features of the proofs grouped into one cluster already exists, as a prototype, see Figures~\ref{automaton} and~\ref{automaton2}. The synthesis of important features from several term-trees is a harder task; the problem is discussed in detail in~\cite{lpar13} -- currently, ML4PG can only show the term tree associated with a statement (cf. Figure~\ref{ml4pgtermtree}).

 Note also that  ML4PG, similarly to dependency graphs of type DG-1 takes into account mutual dependency of tactics and Coq object definitions, but for the post-processing stage, it makes differentiation between features of the object definitions and features of the proofs. Thus, the user can choose to see the analysis of similarities of object definitions (cf. Figure~\ref{ml4pgtermtree}); or the analysis of similarities among proofs of the library (cf. Figures~\ref{automaton} and~\ref{automaton2}). 
This is made to facilitate proof development, as we assume ML4PG most common use would be to use clustering outputs as hints in new proofs; see~\cite{HK14mcs} for a detailed analysis of this user scenario.

\subsection{Dependency graphs showing relations of the (lemmas in the) library}

ML4PG's clustering results can be compared to the dependency graphs of type DPG-2 as follows. Clustering results show similarity, rather than dependency graphs.
Two proofs or terms may use completely different data structures (one e.g. nat -- and another -- list), and still they could be grouped together because they are structurally similar.

\begin{example}
The lemma \verb"dpath_path_l"  (cf. Example~\ref{ex1}) is clustered together with the following lemma:

\begin{lstlisting}
Lemma transport_paths_lr {A : Type} {x1 x2 : A} (p : x1 = x2) 
  (q : x1 = x1) : transport (fun x => x = x) p q = p^ @ q @ p.
Proof.
  destruct p; simpl.
  exact ((concat_1p q)^ @ (concat_p1 (1 @ q))^).
Qed. 
\end{lstlisting}

The statements of these two lemmas are quite different and term-clustering does not group them together (cf. Figure~\ref{clustergraph2}). However, the proofs of both results are really similar  (see Figure~\ref{automaton2}) and for this reason they are clustered together regarding their proofs (cf. Figure~\ref{clustergraph1}).

\begin{figure}
\centering 
\includegraphics[scale=.15]{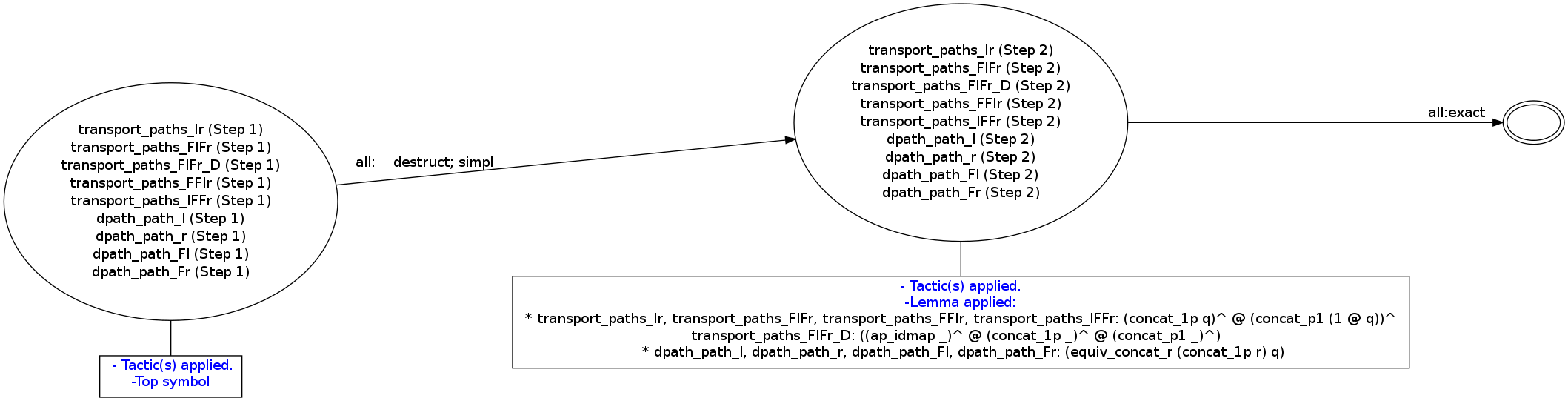} 
\caption{\emph{Automaton for one of the clusters of the Paths library using granularity value 3, showing that the prrofs follows the same strategy (hence single, rather than mutiple, transitions in the automaton.)}}\label{automaton2}
\end{figure}

\end{example}

On the contrary, there can be two lemmas/terms depending on the same term or lemma but being content-wise very different, and then ML4PG will not group them together:

\begin{example}
The statements of lemmas \verb"dpath_path_l" and \verb"dpath_path_lr" are similar and they are grouped in the same cluster using term-clustering (cf. Figure~\ref{clustergraph2}). However, their proofs are different and they are not grouped in the same cluster using proof-clustering (cf. Figure~\ref{clustergraph1}). 
\end{example}

Visualizing results that ML4PG obtains after recurrent library clustering, we get \emph{similarity graphs} as opposed to \emph{dependency graphs}, see Figures~\ref{clustergraph2}  
and~\ref{clustergraph1}.
In those figures, we see how Coq objects are related, based on their structures and recurrent relations to other Coq terms in the library.

\section{Conclusions}

ML4PG is a statistical machine-learning tool that could be used to post-process information available in the dependency graphs

\begin{itemize}
	\item ML4PG is  discriminative (unlike DG-1), and allows to abstract away from ``unimportant features''. Importance of features is decided after clustering makes statistical comparison of
	all objects (definitions and proofs) of the given libraries.
\item ML4PG separates the output for term-structure and proof-structure, for readability and convenience, although it processes term structures, type structures and proof structures in their integrity during the recurrent library clustering.
\item ML4PG can work beyond direct cross-referencing; hence can show similarities that are not captured by dependencies. In the opposite direction, 
it discriminates the information about objects that depend on same structures in case these objects are dissimilar on the ``big picture''.

\end{itemize}

\bibliographystyle{plain}
\bibliography{biblio}

\end{document}